\documentclass[%
preprint,
amsmath,amssymb,
prb,
]{revtex4-2}

\usepackage{graphicx}
\usepackage{dcolumn}
\usepackage{bm}
\usepackage[section]{placeins}
\usepackage[colorlinks=true, citecolor=blue, linkcolor=blue, urlcolor=blue]{hyperref}

\begin{document}

	\title{Recurrence in a periodically driven and weakly damped Fermi-Pasta-Ulam-Tsingou chain}
	
	\author{Yujun Shi}
	\email{yujunshi@sxu.edu.cn}
	\affiliation{College of Physics and Electronic Engineering, Shanxi University, Taiyuan 030006, China}
	
	\author{Haijiang Ren}
	\affiliation{College of Physics and Electronic Engineering, Shanxi University, Taiyuan 030006, China}


	\begin{abstract}
		
		We report numerical evidence of Fermi-Pasta-Ulam-Tsingou (FPUT)–like recurrence in weakly damped, periodically driven $\alpha$-FPUT chains. In narrow regions of driving amplitude and damping, the steady-state energy is exchanged among a few low-frequency modes in a quasi-periodic (or highly regular, near-periodic) manner over long timescales. The maximum damping allowing recurrence decreases rapidly with chain length, suggesting that in the thermodynamic limit such behavior is unlikely to persist. Unlike discrete time crystals, the recurrence period is not an integer multiple of the driving period and does not correspond to spontaneous symmetry breaking. Nevertheless, these results reveal a new type of coherent nonlinear dynamics in driven, open multimode systems and provide guidance for experimentally realizing long-lived quasi-periodic states.

	\end{abstract}
	
	\maketitle
	
	\section {Introduction}\label{sec1}
	
	Fermi–Pasta–Ulam–Tsingou (FPUT) recurrence is a nonlinear phenomenon in multimode systems, where energy is exchanged primarily among a few low-frequency modes and quasi-periodically returns close to its initial distribution, instead of rapidly equi-partitioning across all modes\cite{osti_4376203,Dauxois_2005,10.1063/1.1855036,e662244e-d100-3772-8e3d-16904c229ac5}. As one of the celebrated problems in nonlinear science, it has attracted extensive theoretical and numerical investigation since its original report by Fermi, Pasta, Ulam, and Tsingou in 1953\cite{osti_4376203}. The prevailing understanding today is that the system eventually thermalizes and reaches energy equipartition, albeit over exceedingly long timescales\cite{10.1063/1.3658620,doi:10.1073/pnas.1404397112}. Thus, the recurrence is now viewed as a long-lived metastable regime, can be referred to as a prethermalized state\cite{tg2c-tmx9}. 
	
\begin{figure}[htbp]
		\includegraphics[width =0.75\textwidth]{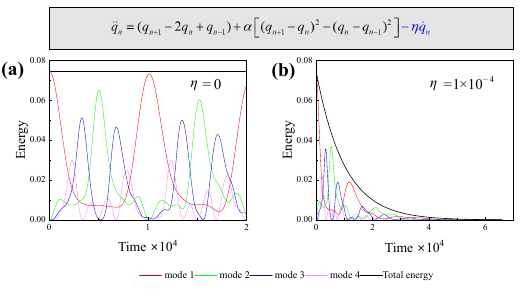}
		\caption{(a) The FPUT recurrence in a chain of length \( N = 32 \) with \(\alpha = 0.25\). 
			(b) Multiple recurrence cycles are suppressed even when a small damping \(\eta = 1\times10^{-4}\) is introduced. }
		\label{Figure01}
	\end{figure}

	Experimentally, observing FPUT recurrence remains highly challenging. To date, only a few reports have demonstrated recurrence-like behavior in various nonlinear systems, such as LC circuits, surface gravity waves, magnetic film-based active feedback rings, optical microresonators, and optical fibers\cite{doi:10.1143/JPSJ.28.1366,PhysRevLett.117.144102,PhysRevLett.98.047202,PhysRevLett.117.163901,PhysRevLett.87.033902,2018NaPho..12..303M,PhysRevX.8.041017}. However, in all these cases, the observed recurrence typically lasts for only a single cycle and occurs no more than three times\cite{PhysRevX.8.041017}. One of the main obstacles is dissipation. FPUT recurrence was originally predicted for idealized Hamiltonian systems with strict energy conservation. In contrast, real physical systems inevitably involve dissipation, which severely limits the possibility of observing multiple recurrence cycles\cite{PhysRevX.4.011054}, as illustrated in Fig.\ref{Figure01}. Consequently, if one adopts the criterion of sustained recurrence over many cycles, it is reasonable to conclude that a genuine long-time FPUT recurrence has yet to be achieved experimentally.
	
	Given that dissipation is unavoidable in most real physical systems, a natural question arises: can external driving compensate for energy loss and enable recurrence-like behavior? Specifically, can a dissipative, multimode nonlinear system--when subject to periodic forcing, exhibit a periodic or near-periodic response over timescales far exceeding the driving period, with energy predominantly exchanged among a small subset of modes, in a manner reminiscent of the original FPUT recurrence?
	
	It is well known that multimode nonlinear systems, involving many degrees of freedom, exhibit a rich variety of dynamical behaviors. As a representative example, consider a one-dimensional chain of anharmonic oscillators under periodic driving. When the driving frequency lies within the spectral band of the system’s linear dispersion relation and the driving amplitude is small, the system behaves similarly to its linear counterpart: primary resonance peaks emerge near the eigenfrequencies of the normal modes, and the response is a steady-state oscillation synchronized with the driving period. As the driving amplitude increases, nonlinear effects become more pronounced, leading to the appearance of secondary resonances in the vicinity of the primary ones. In this regime, the system may exhibit not only periodic responses but also quasi-periodic or even chaotic dynamics\cite{GEIST1988103}. When the driving frequency lies outside the linear spectrum, for instance, near the band edge, the system can display more intricate dynamical phenomena, such as anharmonic waves, solitons, and breathers\cite{PhysRevResearch.3.023106,PhysRevB.50.9652,PhysRevE.64.056606,PhysRevLett.102.205505,PhysRevLett.102.224101,PhysRevE.98.062205}. However, to the best of our knowledge, no FPUT-like recurrence involving multiple modes has been reported in such driven and damped chains. Nonstationary responses, characterized by periodic or quasiperiodic energy exchange, have so far been observed only in two-mode nonlinear systems\cite{nayfeh2024nonlinear,PhysRevLett.52.922,PhysRevLett.123.083901}.
	
	As shown in this work, a key reason for the absence of such observations lies in the magnitude of the damping. In most previous studies of driven, damped anharmonic chains, the damping coefficient was typically chosen to be relatively large--often not smaller than 0.1--to accelerate convergence to a steady state in numerical simulations\cite{GEIST1988103,PhysRevE.64.056606,PhysRevLett.102.205505,PhysRevLett.102.224101}. Our results demonstrate that the emergence of robust recurrence behavior requires much smaller damping.
	
	In this Letter, we present numerical evidence that a recurrence-like phenomenon can emerge in a periodically driven, weakly damped FPUT chain. Specifically, near the first mode, we identify a narrow region in the parameter space of driving amplitude and damping where energy is periodically exchanged among a few modes over long timescales. The maximum damping that allows such recurrence decreases rapidly with system size. For example, it is approximately \( 4.3 \times 10^{-3} \) for \( N = 8 \), and about \( 5 \times 10^{-5} \) for \( N = 32 \). These results suggest that in the thermodynamic limit, such recurrence behavior is unlikely to persist. We interpret this phenomenon as a new type of coherent nonlinear dynamics enabled by a delicate balance between weak driving, weak damping, and the system's modal structure.
	
	We would like to clarify the use of the term recurrence. Strictly speaking, in a driven-dissipative  system, the long-period behavior observed in the modal energies should not be interpreted as recurrence in the Hamiltonian sense--that is, a return to the initial state, since the system no longer evolves on a constant-energy manifold. The long-time dynamics of driven-dissipative  systems is governed by attractors, and the influence of the initial condition decays during the transient stage. Accordingly, the phenomenon reported in this work is more precisely described as a long-period energy modulation or a coherent energy exchange cycle, rather than a genuine recurrence. Nevertheless, in order to emphasize its phenomenological connection with the original FPUT recurrence, we retain the term recurrence throughout the paper, while acknowledging this distinction.
	
	The paper is organized as follows. Section \ref{sec2} introduces the model and the methodology used in our numerical simulations. Section~\ref{sec3} presents the numerical results, including representative recurrence phenomena and the dependence of the system's response on driving frequency, driving amplitude, damping, and chain length. Finally, in Sec.~\ref{sec4}, we first briefly discuss our numerical simulations from the perspective of soliton dynamics in the KdV equation and in the Toda lattice. We then examine the connection between the observed recurrence behavior and the recently explored concept of time crystals. Appendices~\ref{appendixA} and~\ref{appendixB} provide supplementary materials on recurrence under various driving conditions in the $\alpha$-FPUT chains and on recurrence in the $\beta$-FPUT chains, respectively.

	\section{ Model and methodology for our numerical simulations}\label{sec2}
	
	The equations of motion of the forced and damped \(\alpha\)-FPUT chain are 
	\begin{equation}\label{driven-FPUT}
		\begin{aligned}
			\ddot{q}_n=& (q_{n+1}-2q_n+q_{n-1})+\alpha\left[(q_{n+1}-q_n)^2-(q_n-q_{n-1})^2\right]\\
			&-\eta\dot{q}_n+F\cos(\Omega t), \quad\quad (n=1,2,...,N),
		\end{aligned}
	\end{equation}
	where \(\eta \) is the damping coefficient, \(F\) and \(\Omega\) are the amplitude and frequency of the driving force.  The fixed boundary conditions are chosen: \(q_{0}=q_{N+1}=0\).

	We integrate Eq.~(1) using a Runge-Kutta algorithm (verified with both MATLAB's \texttt{ode45} and Python's \texttt{scipy.integrate.solve\_ivp}, yielding consistent results), with initial conditions \(q_n = 0\) and \(\dot{q}_n = 0\). In the original FPUT problem, Runge-Kutta methods are generally not used because they are not symplectic and therefore do not guarantee long-term energy conservation. In our driven–damped model, however, symplectic algorithm is not required. To ensure numerical accuracy, the solver's relative and absolute error tolerances were reduced to \(10^{-8}\) and \(10^{-12}\), respectively, which are significantly smaller than the default settings.
	
	we have monitored the hamiltonian of the undriven, undamped  \(\alpha\)-FPUT as the ``total energy''
	\begin{equation}
		H=\sum_{n=1}^N\frac{p_n^2}{2}+\sum_{n=0}^N\left[\frac{1}{2}\left(q_{n+1}-q_n\right)^2+\frac{\alpha}{3}\left(q_{n+1}-q_n\right)^3\right],
	\end{equation}
	and energies of the individual normal modes
	\begin{equation}
		E_k=\frac{P_k^2+\omega_k^2Q_k^2}{2},
	\end{equation}
	where the normal-mode frequencies are given by
	\begin{equation}
		\omega_k=2\sin\left(\frac{\mathrm{k}\pi}{2(\mathrm{N}+1)}\right),
	\end{equation}
	and the normal-mode coordinates \((Q_k, P_k)\) are related to the lattice coordinates \((q_n, p_n)\) via
	\begin{equation}
		\begin{bmatrix}q_n\\p_n\end{bmatrix}=\sqrt{\frac{2}{N+1}}\sum_{k=1}^{\mathrm{N}}\begin{bmatrix}\mathrm{Q}_k\\\mathrm{P}_k\end{bmatrix}\sin\left(\frac{nk\pi}{\mathrm{N}+1}\right).
	\end{equation}
	
	\begin{figure}[htbp]
		\centering
		\includegraphics[width=0.9\textwidth]{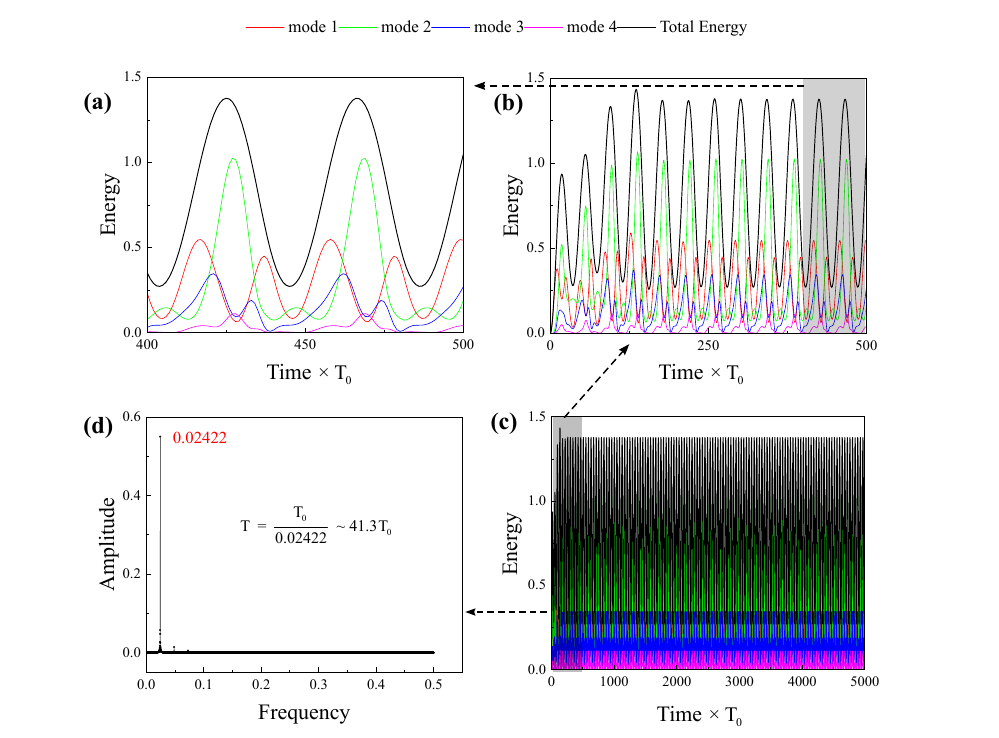}
		\caption{Representative recurrence phenomenon. (a--c) Simulation results for  the \( N = 8 \) chain with \(\eta = 2.5\times10^{-3}\), driving amplitude \(F = 5\times10^{-3}\), and driving frequency \(\omega = 0.347\) (close to the first mode \(\omega_{1} = 0.3473\)). The horizontal axis represents time in units of the driving period \(T_0\). Panel (c) shows the evolution of modal energies over a time span of $5\times10^{3}$ driving periods. Due to the high density of curves, the traces appear as a filled region; this representation is intended solely to demonstrate the long-time persistence of the recurrence phenomenon. Local details are successively magnified in panels (b) and (a). (d) Fast Fourier transform (FFT) of the steady-state response shown in panel (c).
		} 
		\label{Figure02}
	\end{figure}

	\section{ Numerical results}\label{sec3}
	
	Unless otherwise stated, all simulation results presented in this paper correspond to a chain of length \( N = 8 \) with a nonlinear coupling coefficient \(\alpha = 0.25\).
	
	\begin{figure*}
		\centering
		\includegraphics[width=1\textwidth]{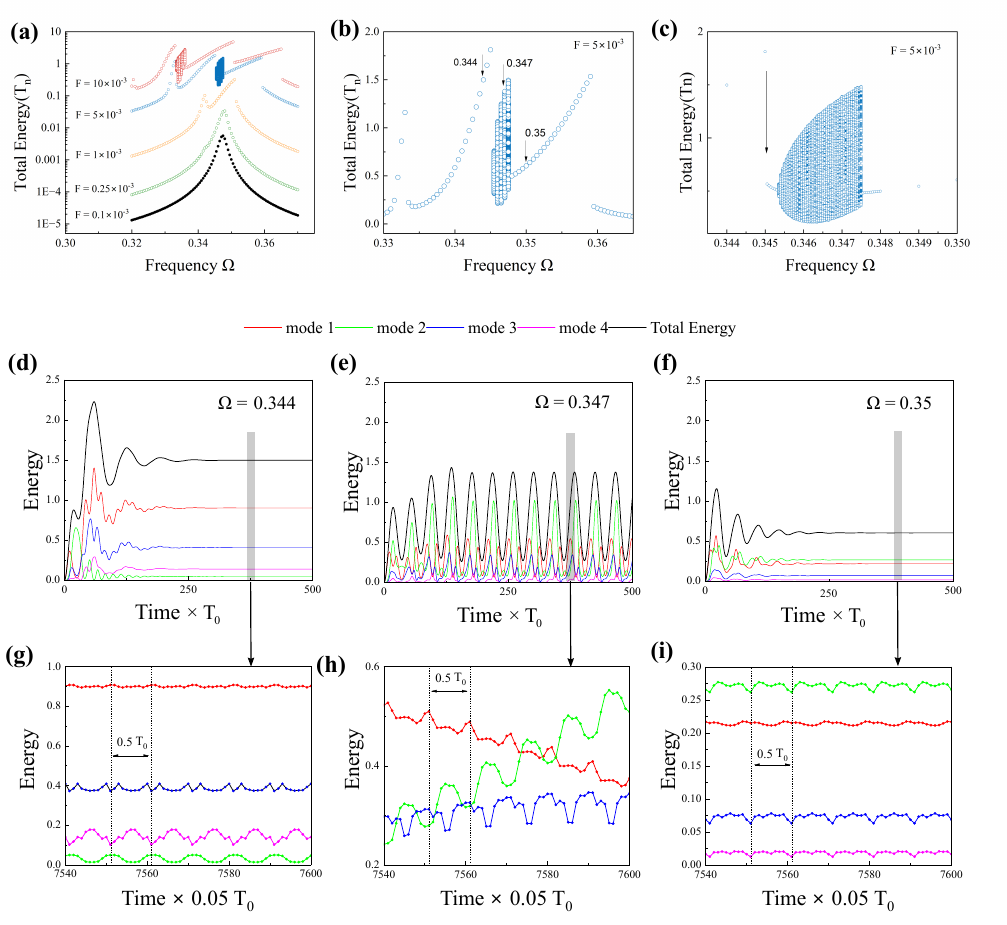}
		\caption{
			(a) Total energy spectra of the \( N = 8 \) chain as a function of the driving frequency near the first mode \(\omega_{1} = 0.3473\), \(\eta = 2.5\times10^{-3}\), for different driving amplitudes. 
			(b) The same data for \( F = 5\times10^{-3} \) shown separately, with a linear scale on the vertical axis. (c) Enlarged view of the frequency range where the recurrence phenomena occur, obtained from a finer frequency scan of the section shown in (b). (d--f) Time-domain simulations for three representative driving frequencies, sampled once per driving period $T_0$. (g--i) Enlarged views of the shaded regions in panels (d--f), sampled at intervals of $0.05\,T_0$. In panels (g--i), the energies of all modes exhibit oscillations with a period of $0.5\,T_0$, indicating that all modes oscillate at the driving frequency.
		}
		\label{Figure03}
	\end{figure*}
	
	\subsection{ Recurrence behavior at \(\eta = 2.5\times10^{-3}\)}
	
	\subsubsection{Representative recurrence behavior  }
	
	Figure \ref{Figure02} shows a representative recurrence behavior for a driving amplitude \( F = 5\times10^{-3} \) and a driving frequency \(\omega = 0.347 \) (near the first mode \(\omega_{1} = 0.3473 \)). 
	Instead of reaching a steady response, there is a continual exchange of energy between the low-frequency modes, exhibiting an almost periodic behavior with a period of approximately \(41.3~T_{0}\) (\(T_{0}\) is the driving period).

	\subsubsection{Energy spectra versus driving frequency}
	
	Figure \ref{Figure03}(a) shows the total energy response as a function of the driving frequency near the first mode $\omega_{1} = 0.3473$, for different driving amplitudes. Specifically, for each set of control parameters, Eq.~\ref{driven-FPUT} was integrated for $N$ driving periods $T_0$ (typically $N = 2000$) until transients had decayed. The total energy was then stroboscopically sampled with period $T_0$ over the following $500\,T_0$, and these data points are plotted in Fig.~\ref{Figure03}(a). To maintain consistency, the first sampled point for each frequency scan was taken at the maximum total energy within one $T_0$ period.
	
	As the driving amplitude increases, nonlinear effects become more pronounced, giving rise to secondary resonances near the primary peaks, accompanied by jump phenomena in their vicinity. The recurrence phenomena occur in the vicinity of these jumps, which is more clearly seen in Fig.~\ref{Figure03}(b) and (c). On both sides of the recurrence regions, the system still exhibits normal steady-state oscillations synchronized with the driving period (Figs.~\ref{Figure03}(d) and (f)). It is worth noting that, regardless of whether recurrence occurs, the system still oscillates at the driving frequency. In particular, every individual mode of the system oscillates at the driving frequency, as shown in Fig.~\ref{Figure03}(g–i).
	
\begin{figure}[htbp]
		\centering
		\includegraphics[width=0.45\textwidth]{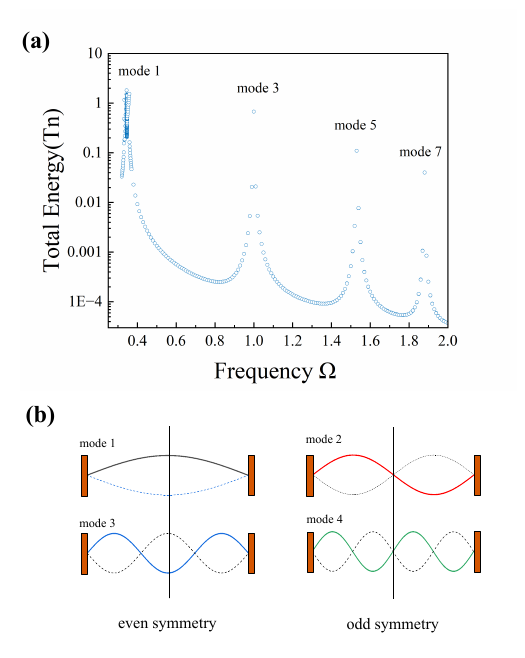}
		\caption{ (a) Total energy spectra of the \( N = 8 \) chain as a function of the driving frequency from 0.32 to 2, for the driving amplitudes \( F = 5\times10^{-3} \). Because the chain is driven uniformly, only modes with even spatial symmetry about the chain center are excited, so the primary resonance peaks appear only near the odd-numbered mode frequencies. (b) Spatial symmetry of the normal modes.}
		\label{Figure04}
	\end{figure}

	A closer look at the recurrence phenomenon at $\Omega = 0.347$, particularly in Fig.~\ref{Figure02}(a), reveals that the average energy of mode 2 exceeds that of mode 1. Given that $\omega_2 = 0.684 \sim 2 \omega_1$, one may speculate that this recurrence arises because the driving at $\Omega = 0.347$ simultaneously excites mode 1 (primary resonance) and mode 2 (secondary, specifically a super-harmonic resonance). The phase mismatch and mutual coupling between these modes lead to periodic energy exchange among them. We can exclude this possibility that the driving simultaneously injects energy into both mode 1 and mode 2. This is because our uniform driving field is even symmetry with respect to the chain center; consequently, whether it is the primary or secondary resonance, only modes with the same spatial symmetry can be directly excited. Direct energy injection into odd-symmetric even-numbered modes is forbidden. This is illustrated in Fig.~\ref{Figure04}(a), which shows the total energy spectra over a wider driving frequency range (from 0.32 to 2) with $F = 5\times10^{-3}$. Only the primary resonances corresponding to odd-numbered modes are excited. Therefore, the only pathway for energy injection is through the initial excitation of mode 1, followed by transfer to higher-order modes via nonlinear coupling, ultimately resulting in a recurrence characterized by a dynamic balance among the low-frequency modes.  

\begin{figure}[htbp]
		\centering
		\includegraphics[width=1\textwidth]{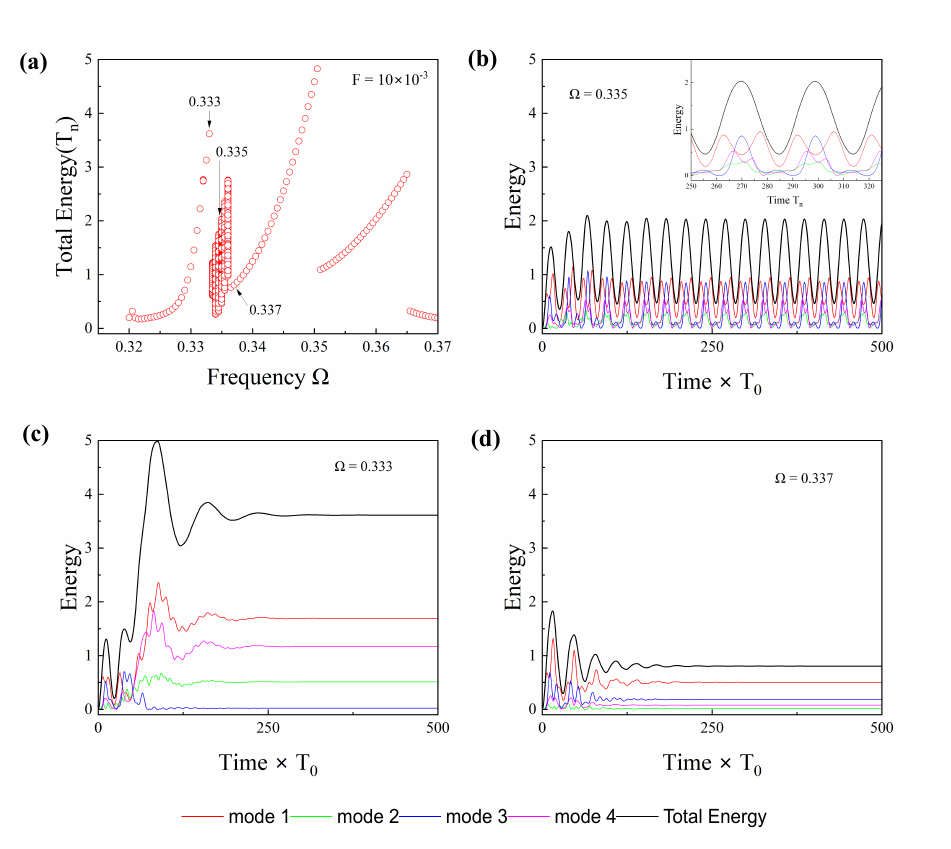}
		\caption{(a) Total energy spectra of the \( N = 8 \) chain as a function of the driving frequency near the first mode \(\omega_{1} = 0.3473\), for  driving amplitudes \( F = 10\times10^{-3} \). (b--d) Time-domain simulations for three representative driving frequencies.}
		\label{Figure05}
	\end{figure}

	Figure~\ref{Figure05} shows the energy response for \( F = 10\times10^{-3} \), which corresponds to one of the data traces in Fig.~\ref{Figure03}(a) plotted separately for clarity. The recurrence phenomenon at \( \Omega = 0.335 \) [see Fig.~\ref{Figure03}(b)] indicates that the energy is mainly concentrated in modes~1 and~3, in contrast to the case of \( F = 5\times10^{-3} \), where modes~1 and~2 dominate.
	
	In addition to uniform driving, we also examined three other types of excitation: 
		(i) single-site driving,
		\( F_n = a\,\delta_{n,n_0}\cos(\Omega t) \)\cite{GEIST1988103};
		(ii) mode-matched driving,
		\( F_n = a\,\sin\!\left(\tfrac{n k \pi}{N+1}\right)\cos(\Omega t) \); 
		and (iii) staggered driving,
		\( F_n = a\,(-1)^n\cos(\Omega t) \),
		which models a uniform electric field applied to a chain of alternating charges\cite{PhysRevE.64.056606}. Both cases produce similar recurrence behaviors, as shown in the Appendix \ref{appendixA}, indicating that the recurrence is an intrinsic consequence of the system’s nonlinear mode coupling rather than a specific feature of the excitation.

	\subsubsection{Energy spectra versus driving amplitude}
	
\begin{figure}[htbp]
		\centering
		\includegraphics[width=0.75\textwidth]{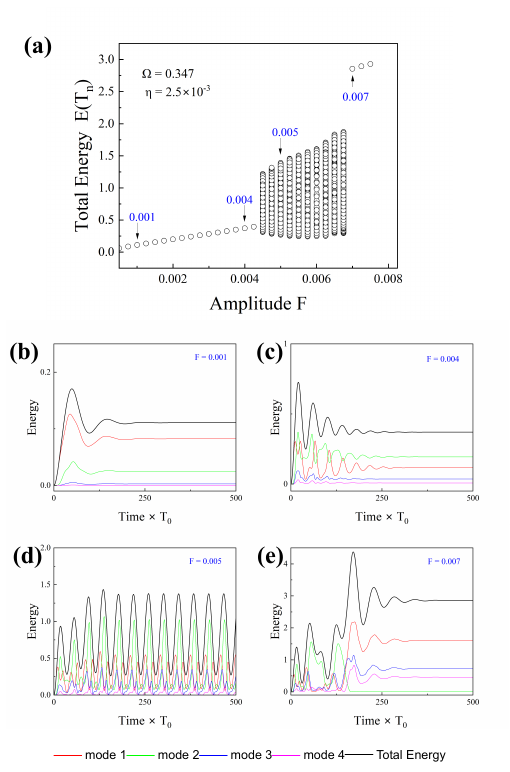}
		\caption{(a) Total energy spectra of the \( N = 8 \) chain as a function of the driving amplitude at the driving frequency \(\Omega = 0.347\). 
			(b–e) Time-domain simulations for four representative driving frequencies.}
		\label{Figure06}
	\end{figure}

	To further examine how the driving amplitude influences the excitation dynamics, an amplitude scan was performed at \( \Omega = 0.347 \) [Fig.~\ref{Figure06}]. 
	At low amplitudes (\( F = 1\times10^{-3} \)), the system exhibits normal steady-state oscillations dominated by mode~1. 
	With increasing amplitude, enhanced nonlinear coupling between modes~1 and~2 causes the energy to shift gradually toward mode~2, which becomes dominant in the recurrence region. 
	Beyond this region, the system resumes normal steady-state oscillations, again dominated by mode~1.

	\begin{figure}[htbp]
		\centering
		\includegraphics[width=0.75\textwidth]{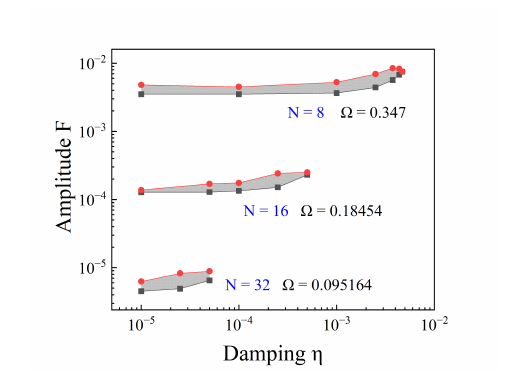}
		\caption{Recurrence regions in the driving amplitude–damping plane for chains of length \(N = 8\), \(16\), and \(32\), each driven at a frequency close to its first normal mode. Red filled circles and black filled squares indicate the boundary points of each region, and the connecting lines are drawn only to guide the eye. Note the logarithmic scale: the parameter space in which recurrence is observed becomes significantly narrower for \(N = 32\) compared to \(N = 8\). The recurrence regions have an apparent lower boundary at \(\eta = 10^{-5}\) due to the artificial limit on the damping; smaller values would require much longer simulations to reach a steady state.		
		}
		\label{Figure07}
	\end{figure}
	
	\subsection{Recurrence regions versus chain Length}
	
	Finally, we investigated how the recurrence regions depend on the chain length. 
	Given many parameters in the model, we fixed the nonlinear coupling coefficient at \( \alpha = 0.25 \) and performed scans in the driving amplitude–damping plane separately for chains of length \( N = 8, 16, \) and \( 32 \), using a fixed driving frequency near mode~1, respectively.
	As shown in Fig.~\ref{Figure07}, the parameter space in which recurrence is observed becomes significantly narrower with increasing chain length. For \( N = 32 \), the maximum damping that allows such recurrence is about \( 5 \times 10^{-5} \), which already represents a very stringent requirement for realistic physical systems. This suggests that, in the thermodynamic limit (\( N \to \infty \)), such recurrence behavior is unlikely to persist.

	\section{Discussion and Conclusion }\label{sec4}
	
We report numerical evidence of recurrence phenomena in driven--damped $\alpha$-FPUT chains. We also performed similar investigations for driven--damped $\beta$-FPUT chains. For a chain of length $N=32$ with $\beta = 8$, analogous recurrence was observed (see Appendix~\ref{appendixB} for details). The choice of $N=32$ and $\beta = 8$ is consistent with the parameter settings used in the original work of Fermi \emph{et al.}~\cite{osti_4376203}. 
		
	Moreover, for the undriven and undamped FPUT system, recurrence has been famously related to soliton dynamics in the KdV equation or in the Toda lattice\cite{PhysRevLett.15.240,doi:10.1143/JPSJ.22.431,10.1063/1.5122972}. Both the KdV equation and the Toda lattice are integrable models that admit analytical descriptions. In contrast, the system studied here is a finite-length FPUT chain subject to both external driving and damping, a setting that fundamentally breaks integrability. In particular, the reduction of the FPUT lattice to the KdV equation is only valid under a continuum approximation with specific weakly nonlinear, long-wavelength, and unidirectional asymptotic scalings. As a result, the soliton-based interpretation of recurrence may not directly applicable to the present system. From a theoretical perspective, the most feasible analytical route would be a perturbative treatmen\cite{10.1063/1.1703904}. However, even for a short chain with $N=8$, such an approach becomes mathematically cumbersome and technically demanding due to the presence of both driving and damping. Consequently, deriving analytical approximations that describe the system's behavior inside the recurrence regions remains challenging, providing a natural direction for future theoretical exploration.
	
	Interestingly, this long-time recurrence bears a superficial resemblance to phenomena recently studied in the context of discrete time crystals (DTCs)\cite{10.1063/PT.3.4020,PhysRevLett.123.124301,Yao2020}. The periodic drive defines a discrete time-translation symmetry group generated by the operation 
	\(\mathcal{T}_T : t \mapsto t + T\).
	A discrete time crystal corresponds to a spontaneous breaking of this symmetry, where the system’s response becomes periodic with an integer multiple of the driving period, i.e., a subgroup of the original time-translation group. In contrast, the long-period recurrence observed in our driven FPUT chain does not represent such symmetry breaking, because its recurrence period is not an integer multiple of the driving period, and the corresponding operation does not form a subgroup of the original symmetry group defined by the drive.
	
	Nonetheless, this comparison highlights an intriguing analogy. The recurrence windows identified here can be regarded as a generalized, relaxed form of time-crystalline order, in which the system exhibits long-term temporal organization and approximate subharmonic responses. Understanding whether such structured recurrence can persist or be stabilized under modified conditions, such as weak noise, feedback control, or quasi-periodic driving, could provide a new bridge between classical nonlinear lattice dynamics and the emerging physics of time crystals, suggesting that classical multimode recurrence phenomena may serve as a testbed for exploring time-crystalline-like behavior.
	
	Finally, returning to the original motivation of this work, our results indicate that FPUT recurrence phenomena, traditionally studied in isolated Hamiltonian systems, can also manifest in driven, open systems accessible in laboratory experiments.

	\section*{Acknowledgments}
	This research was supported by the Research Project Supported by Shanxi Scholarship Council of China (Grant No. 2023-033), and the Fundamental Research Program of Shanxi Province (Grant No. 202303021221071).
	

	\appendix
	\section{Recurrence under Various Driving Conditions}\label{appendixA}
	
\begin{figure}[htbp]
		\centering
		\includegraphics[width=0.75\textwidth]{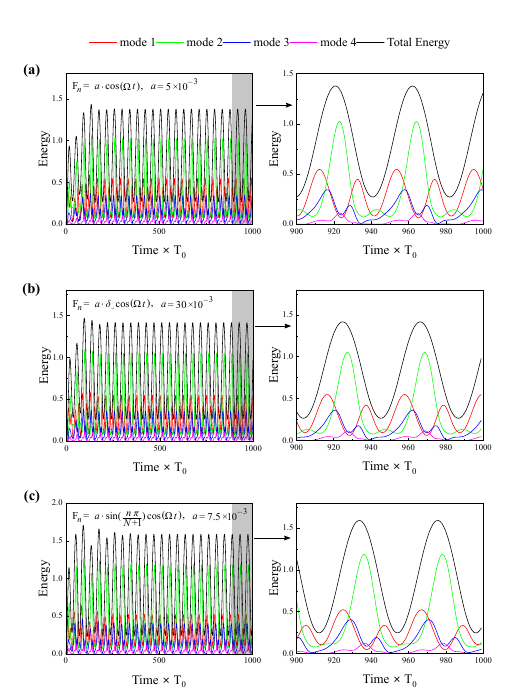}
		\caption{Recurrence behavior of an $N=8$ FPUT chain under uniform, single-site, and mode 1-matched driving at $\Omega = 0.347$, with damping $\eta = 2.5\times10^{-3}$ and nonlinear coupling coefficient $\alpha = 0.25$.
		}
		\label{Figure08}
	\end{figure}

\begin{figure}[htbp]
		\centering
		\includegraphics[width=0.6\textwidth]{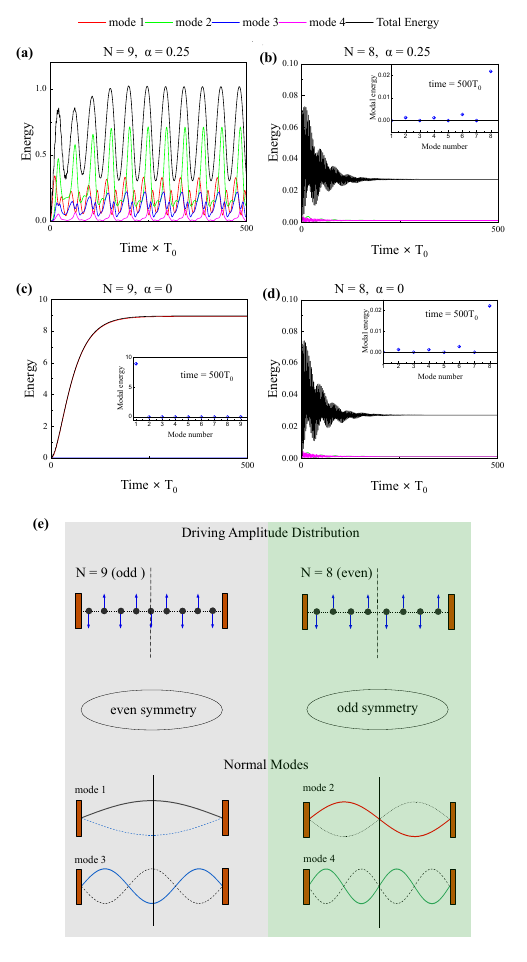}
		\caption{
			(a) and (b) Staggered driving, $F_n = a\,(-1)^n \cos(\Omega t)$, for $N=9$ (odd, recurrence) and $N=8$ (even, no recurrence), driven near the first normal mode, $\Omega = 0.347$ and $0.313$, respectively, with $\eta = 2.5\times10^{-3}$ and $\alpha = 0.25$. 
			(c) and (d) show the excitation behavior of the linear chain ($\alpha = 0$): mode~1 dominates for $N=9$, while odd modes are forbidden for $N=8$. The insets in panels (b--d) display the energy distribution among the normal modes at time $t = 500\mathrm{T_0}$. Note that mode~8 eventually dominates for $N=8$ because it is the best spatial match to the staggered driving. (e) The spatial symmetry of the driving amplitude distribution and the normal modes.
		}
		\label{Figure09}
	\end{figure}

	Figure~\ref{Figure08} shows the recurrence behavior of an $N=8$ FPUT chain under uniform, single-site, and mode 1-matched driving at $\omega = 0.347$. 
	Figure~\ref{Figure09} presents results for FPUT chains under a staggered driving, 
	\(F_n = a\,(-1)^n \cos(\Omega t)\), with the same driving amplitude $F = 0.15$ and damping $\eta = 2.5\times10^{-3}$.
	For staggered driving, the site-dependent driving amplitude distribution exhibits different symmetries with respect to the chain center, depending on the chain length: it is even-symmetric for odd $N$ and odd-symmetric for even $N$ (Fig.~\ref{Figure09}(e)). 
	Consequently, when driving near the frequency of the first normal mode, which has even symmetry, direct excitation of mode~1 is possible only for chains with odd length $N$ (see Fig.~\ref{Figure09}(c--d), where the nonlinear coupling coefficient is set to zero, $\alpha = 0$).
	Figures~\ref{Figure09}(a) and~\ref{Figure09}(b) correspond to driving near the first normal mode for $N=9$ and $N=8$, respectively. 
	A recurrence phenomenon similar to that observed under uniform driving is clearly present for the odd-length chain, whereas no such recurrence is observed for the even-length chain. In fact, staggered driving is highly inefficient for chains with even $N$. As shown in Fig.~\ref{Figure09}(b), the steady-state total energy reaches only about $0.027$, which is much smaller than that in Fig.~\ref{Figure09}(a) and is almost indistinguishable from the linear-chain response in Fig.~\ref{Figure09}(d). This indicates that the system remains in a weakly excited regime where nonlinear coupling plays a negligible role.

	\section{Recurrence in the driven damped \(\beta\)-FPUT chains}\label{appendixB}
\begin{figure}[htbp]
		\centering
		\includegraphics[width=0.85\textwidth]{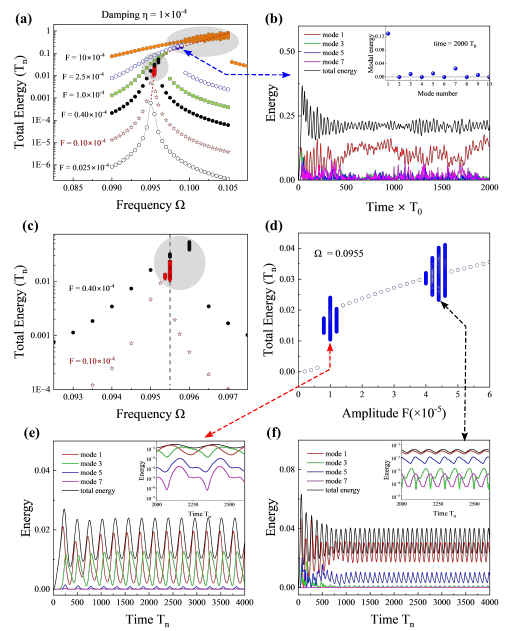}
		\caption{
			Uniform driving (\( F_n = F\cos(\Omega t) \)) with $N = 32$, $\beta = 8$, and $\eta = 10^{-4}$. 
			(a) Total-energy spectra as a function of the driving frequency near the first normal-mode frequency, $\omega_1 = 0.09516$, for different driving amplitudes. 
			(b) Time-domain simulations corresponding to the multivalued total energy in panel (a) ($\Omega = 0.099$, $F = 2.5~\times 10^{-4}$), illustrating a chaotic steady state. 
			(c) Enlarged view of panel (a) showing the local structure.
			(d) Total-energy spectra as a function of the driving amplitude at fixed driving frequency $\Omega = 0.0955$. 
			(e–f) Recurrence phenomena observed in the two multivalued regions in panel (d). In the insets, the vertical axis is plotted on a logarithmic scale to facilitate comparison of the modal energy magnitudes.
		}
		\label{Figure10}
	\end{figure}

\begin{figure}[htbp]
		\centering
		\includegraphics[width=0.85\textwidth]{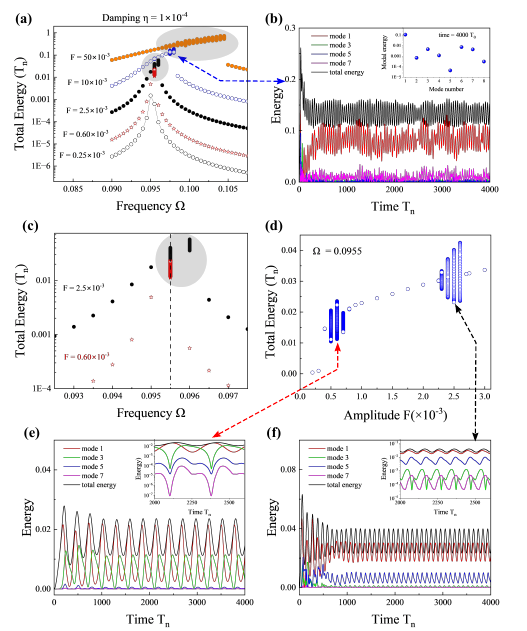}
		\caption{
		Single site driving (\( F_n = F\,\delta_{n,4}\cos(\Omega t) \)) with $N = 32$, $\beta = 8$, and $\eta = 10^{-4}$. 
		(a) Total-energy spectra as a function of the driving frequency near the first normal-mode frequency, $\omega_1 = 0.09516$, for different driving amplitudes. 
		(b) Time-domain simulations corresponding to the multivalued total energy in panel (a) ($\Omega = 0.0975$, $F = 10~\times 10^{-3}$), illustrating a chaotic steady state. 
		(c) Enlarged view of panel (a) showing the local structure.
		(d) Total-energy spectra as a function of the driving amplitude at fixed driving frequency $\Omega = 0.0955$. 
		(e–f) Recurrence phenomena observed in the two multivalued regions in panel (d). In the insets, the vertical axis is plotted on a logarithmic scale to facilitate comparison of the modal energy magnitudes.
	}
		\label{Figure11}
	\end{figure}

	For a chain of length $N=32$ with $\beta = 8$, we performed scans of the total-energy spectra as functions of the driving frequency in the vicinity of the first normal-mode frequency ($\omega_1 = 0.09516$). The scans were carried out under both uniform driving (symmetric under lattice reflection, where excitation of even modes is forbidden) and single-site driving (non-symmetric under lattice reflection, where excitation of even modes is allowed). Within the explored parameter ranges, analogous recurrence behavior was observed (see Figs.~\ref{Figure10} and \ref{Figure11} for details).
	
	It can be seen that Figs.~\ref{Figure10} and \ref{Figure11} are nearly identical, despite corresponding to different driving parameters. This indicates that the recurrence phenomenon is robust to different forms of the driving term. When the system reaches comparable total energies, the recurrence patterns under different driving forms are nearly the same.
	
	Furthermore, in both driving protocols of the $\beta$-FPUT chain, we find that the system's energy is primarily concentrated in the four odd modes 1, 3, 5, and 7. Under uniform driving, the energies of the even modes remain strictly zero. Under single-site driving, the even modes are excited, but their energies are significantly smaller in magnitude compared with the four dominant odd modes.
	
	To illustrate this point, in the normal-mode coordinates, the equations of motion can be written as follows:
	
	\begin{equation}
		\ddot{Q}_{k}=-\omega_{k}{}^{2}Q_{k}-\frac{\beta}{4}\sum_{i,j,m=1}^{N}B_{k, ijm}Q_{i}Q_{j}Q_{m}+f_{k}(\Omega t)+\sum_{l}^{N}\eta_{l}\dot{Q}_{l}
	\end{equation}
	where the term $f_k(\Omega t)$ corresponds to the external driving, 
	$\eta_l \dot{Q}_l$ represents dissipation, 
	and the mode-coupling coefficients are given by:
	\begin{equation}\label{B_coupling}
		B_{k, ijm}=\frac{\omega_k\omega_i\omega_j\omega_m}{2(N+1)}\sum_{\pm}\left[\delta_{k,\pm i\pm j\pm m}-\delta_{k\pm i\pm j\pm m,\pm2(N+1)}\right]
	\end{equation}
	
	Since the periodic driving frequency is close to the first normal-mode frequency, 
	if mode coupling is neglected, the energy injected directly by the external driving 
	is concentrated primarily in mode 1, while direct excitation of the other modes 
	diminishes with increasing frequency detuning. Analogous to the original (undriven) FPUT problem, this situation is, to some extent, equivalent to having the initial energy highly concentrated in mode 1,
	that is, the initial values satisfy $Q_{k\neq 1} \ll Q_1$ and $\dot{Q}_k = 0$. 
	According to Eq.~\ref{B_coupling}, a necessary condition for the mode-coupling coefficient 
	$B_{\text{even},ijk}$ to be nonzero is that at least one of $i$, $j$, or $k$ is even.
	Therefore, the corresponding coupling term
	\[
	B_{\text{even}, ijm} Q_i Q_j Q_m
	\]
	remains small, since at least one of the coordinates $Q_i$, $Q_j$, or $Q_m$ is small.
	As a consequence, the coordinate amplitudes of the even modes remain negligibly small throughout the evolution.
	
	In contrast, the first few odd modes can gradually acquire a finite fraction of the total energy through the coupling terms
	\[
	B_{3,111}\, Q_1 Q_1 Q_1, \quad 
	B_{5,113}\, Q_1 Q_1 Q_3, \quad \text{and} \quad 
	B_{7,115}\, Q_1 Q_1 Q_5,
	\]  
	allowing these modes to progressively share energy during the system's evolution.
	
%
	
	\FloatBarrier	
	\bibliography{reference}
	
\end{document}